\documentclass[journal]{IEEEtran}
\usepackage{amsmath,amsfonts}
\usepackage{algorithmic}
\usepackage{array}
\usepackage[caption=false,font=normalsize,labelfont=sf,textfont=sf]{subfig}
\usepackage{textcomp}
\usepackage{stfloats}
\usepackage{url}
\usepackage{verbatim}
\usepackage{graphicx}
\usepackage{soul}

\usepackage{xspace}
\xspaceaddexceptions{-}
\usepackage{glossaries}
\usepackage{color}
\usepackage[hidelinks]{hyperref}

\usepackage{graphicx}        
\graphicspath{{./figures/}}

\usepackage{cite}

\let\OLDthebibliography\thebibliography
\renewcommand\thebibliography[1]{
	\OLDthebibliography{#1}
	\setlength{\parskip}{0pt}
	\setlength{\itemsep}{0pt}
}

\usepackage{siunitx}
\usepackage{amsmath,amssymb,latexsym}
\usepackage{amsfonts}
\usepackage{upgreek}
\usepackage{mathtools}
\usepackage{nicefrac}

\usepackage{multirow}
\usepackage[table,xcdraw]{xcolor}
\usepackage{array}
\newcolumntype{P}[1]{>{\centering\arraybackslash}p{#1}}
\usepackage{booktabs} 
\usepackage[flushleft]{threeparttable} 

\newacronym{ASIen}{ASI}{Italian Space Agency}
    
\newacronym{ASIit}{ASI}{Agenzia Spaziale Italiana}
    
\newacronym{UNICALen}{UNICAL}{University of Calabria}
    
\newacronym{UNICALit}{UNICAL}{Università della Calabria}
    
\newacronym{DIMESen}{DIMES}{Department of Computer Engineering, Modeling, Electronics and Systems}
    
\newacronym{DIMESit}{DIMES}{Dipartimento di Ingegneria Informatica, Modellistica, Elettronica e Sistemistica}
    
\newacronym{USFQes}{USFQ}{Universidad San Francisco de Quito}
         
\newacronym{BIUen}{BIU}{Bar-Ilan University}
     
\newacronym{beol}{BEOL}{back-end-of-line}

\newacronym{feol}{FEOL}{front-end-of-line}

\newacronym{wordl}{WL}{wordline}
    
\newacronym{bitl}{BL}{bitline}
    
\newacronym{sourcel}{SL}{sourceline}
    
\newacronym[longplural={nanowires}]{nw}{NW}{nanowire}

\newacronym{ptm}{PTM}{predictive technology model}

\newacronym{pdk}{PDK}{process design kit}

\newacronym[longplural={systems-on-chip}]{soc}{SoC}{system-on-chip}

\newacronym[longplural={integrated circuits}]{ic}{IC}{integrated circuit}

\newacronym{ai}{AI}{artificial-intelligence}

\newacronym{iot}{IoT}{internet-of-things}

\newacronym{mc}{MC}{Monte Carlo}
     
\newacronym{cvs}{CVS}{conventional voltage sensing}

\newacronym{epi}{EPI}{energy per instruction}

\newacronym{ips}{IPS}{instructions per second}

\newacronym{mep}{MEP}{minimum energy point}

\newacronym{lrs}{LRS}{low resistance state}

\newacronym{hrs}{HRS}{high resistance state}

\newacronym{mim}{MIM}{metal-insulator-metal}

\newacronym[longplural={phase-change memories}]{pcm}{PCM}{phase-change memory}

\newacronym[longplural={resistive RAMs}]{rram}{RRAM}{resistive RAM}

\newacronym[longplural={spin-transfer torque magnetic random-access memories}]{sttmram}{STT-MRAM}{spin-transfer torque magnetic random-access memory}

\newacronym{euv}{EUV}{extreme ultra-violet}
     
\newacronym[longplural={Gain-Cell embedded DRAMs}]{gcedram}{GC-eDRAM}{Gain-Cell embedded DRAM}

\newacronym{sixt}{6T}{6-transistor}
    
\newacronym{eflash}{eFlash}{embedded Flash}

\newacronym[longplural={multi-level cells}]{mlc}{MLC}{multi-level cell}

\newacronym[longplural={Storage Class Memories}]{scm}{SCM}{Storage Class Memory}

\newacronym{ddr}{DDR}{dual-data rate}

\newacronym[longplural={graphic processing units}]{gpu}{GPU}{graphic processing unit}

\newacronym[longplural={central processing units}]{cpu}{CPU}{central processing unit}

\newacronym{sata}{SATA}{Serial Advanced Technology Attachment}
    
\newacronym{nvme}{NVMe}{Non-Volatile Memory Express}
    
\newacronym[longplural={non-volatile memories}]{nvm}{NVM}{non-volatile memory}

\newacronym{pcie}{PCIe}{Peripheral Component Interconnect Express}
     
\newacronym[longplural={hard-Disk drives}]{hdd}{HDD}{hard-Disk drive}

\newacronym[longplural={solid-State drives}]{ssd}{SSD}{solid-State drive}

\newacronym[longplural={high-bandwidth memories}]{hbm}{HBM}{high-bandwidth memory}

\newacronym[longplural={dual-inline memory modules}]{dimm}{DIMM}{dual-inline memory module}

\newacronym[longplural={static random-access memories}]{sram}{SRAM}{static random-access memory}

\newacronym[longplural={embedded DRAMs}]{edram}{eDRAM}{embedded DRAM}

\newacronym[longplural={dynamic random-access memories}]{dram}{DRAM}{dynamic random-access memory}

\newacronym[longplural={six-transistor static random access memories}]{sixtsram}{6T-SRAM}{six-transistor static random access memory}

\newacronym[longplural={magnetic random-access memories}]{mram}{MRAM}{magnetic random-access memory}

\newacronym{bc}{BC}{bitcell}

\newacronym{bl}{BL}{bitline}

\newacronym{sln}{SL}{sourceline}

\newacronym{wl}{WL}{wordline}

\newacronym{nc}{NC}{number of cycles at endurance failure}
     
\newacronym{llgs}{LLGS}{Landau-Lifshitz-Gilbert-Slonczewski}
     
\newacronym{stt}{STT}{spin-transfer torque}

\newacronym{pma}{PMA}{perpendicular magnetic anisotropy}

\newacronym{ima}{IMA}{in-plane magnetic anisotropy}

\newacronym{mtj}{MTJ}{magnetic tunnel junction}

\newacronym{smtj}{SMTJ}{single-barrier MTJ}

\newacronym{dmtj}{DMTJ}{double-barrier MTJ}

\newacronym{mr}{MR}{magnetoresistance}

\newacronym{tmr}{TMR}{tunnel magnetoresistance}

\newacronym{gmr}{GMR}{giant magnetoresistance}

\newacronym{wer}{WER}{write error rate}

\newacronym{rdr}{RDR}{read disturbance rate}

\newacronym{rfr}{RFR}{retention failure rate}

\newacronym{rer}{RER}{read error rate}

\newacronym{fl}{FL}{free layer}

\newacronym[longplural={reference layers}]{rl}{RL}{reference layer}

\newacronym{p}{P}{parallel}

\newacronym{ap}{AP}{antiparallel}

\newacronym{fm}{FM}{ferromagnetic}

\renewcommand{\eqref}[1]{(\ref{#1})}

\definecolor{applegreen}{rgb}{0.55, 0.71, 0.0}

\usepackage{scalerel}
\usepackage{tikz}
\usetikzlibrary{svg.path}
\definecolor{orcidlogocol}{HTML}{A6CE39}
\tikzset{
  orcidlogo/.pic={
    \fill[orcidlogocol] svg{M256,128c0,70.7-57.3,128-128,128C57.3,256,0,198.7,0,128C0,57.3,57.3,0,128,0C198.7,0,256,57.3,256,128z};
    \fill[white] svg{M86.3,186.2H70.9V79.1h15.4v48.4V186.2z}
                 svg{M108.9,79.1h41.6c39.6,0,57,28.3,57,53.6c0,27.5-21.5,53.6-56.8,53.6h-41.8V79.1z M124.3,172.4h24.5c34.9,0,42.9-26.5,42.9-39.7c0-21.5-13.7-39.7-43.7-39.7h-23.7V172.4z}
                 svg{M88.7,56.8c0,5.5-4.5,10.1-10.1,10.1c-5.6,0-10.1-4.6-10.1-10.1c0-5.6,4.5-10.1,10.1-10.1C84.2,46.7,88.7,51.3,88.7,56.8z};
  }
}

\newcommand\orcidicon[1]{\href{https://orcid.org/#1}{\mbox{\scalerel*{
\begin{tikzpicture}[yscale=-1,transform shape]
\pic{orcidlogo};
\end{tikzpicture}
}{|}}}}

\newcommand{\bit}{\,\si{\bit}\xspace}

\def\BibTeX{{\rm B\kern-.05em{\sc i\kern-.025em b}\kern-.08em
    T\kern-.1667em\lower.7ex\hbox{E}\kern-.125emX}}
\usepackage{balance}
\begin{document}
\title{Temperature-Resilient Analog Neuromorphic Chip in Single-Polysilicon CMOS Technology}
\author{
        Tommaso~Rizzo\,\orcidicon{0000-0003-2527-713X},~\IEEEmembership{Member,~IEEE,}
        Sebastiano~Strangio\,\orcidicon{0000-0002-6984-1137},~\IEEEmembership{Senior Member,~IEEE,}
        Alessandro~Catania\,\orcidicon{0000-0001-7242-6228},~\IEEEmembership{Senior Member,~IEEE,}
        and Giuseppe~Iannaccone\,\orcidicon{0000-0003-3375-1647},~\IEEEmembership{Fellow,~IEEE}
\thanks{
{\bf This work has been submitted to the IEEE for possible publication. Copyright may be transferred without notice, after which this version may no longer be accessible.}

Tommaso Rizzo is with Quantavis s.r.l., Largo Padre Renzo Spadoni, 56126 Pisa, Italy and with the Dipartimento di Ingegneria dell'Informazione (DII), Università di Pisa, 56122 Pisa, Italy;
Sebastiano Strangio is with the Dipartimento di Ingegneria dell'Informazione (DII), Università di Pisa, 56122 Pisa, Italy;
Alessandro Catania is with the Dipartimento di Ingegneria dell'Informazione (DII), Università di Pisa, 56122 Pisa, Italy;
Giuseppe Iannaccone is with the Dipartimento di Ingegneria dell'Informazione (DII), Università di Pisa and with Quantavis s.r.l., Largo Padre Renzo Spadoni, 56126 Pisa, Italy (corresponding author:
(e-mail: giuseppe.iannaccone@unipi.it)

This work was partially supported by the EC Horizon 2020 Research and Innovation Programme under GA QUEFORMAL 829035 and under GA AUTOCAPSULE 952118, and by the italian MUR under the Forelab project of the ``Dipartimenti di Eccellenza'' programme.}
}

\markboth{IEEE TRANSACTIONS ON CIRCUITS AND SYSTEMS I -- REGULAR PAPERS,~Vol.~??, No.~??,~??~202?}%
{How to Use the IEEEtran \LaTeX \ Templates}

\maketitle

\begin{abstract}
In analog neuromorphic chips, designers can embed computing primitives in the intrinsic physical properties of devices and circuits, heavily reducing device count and energy consumption, and enabling high parallelism, because all devices are computing simultaneously.  Neural network parameters can be stored in local analog non-volatile memories (NVMs), saving the energy required to move data between memory and logic. However, the main drawback of analog sub-threshold electronic circuits is their dramatic temperature sensitivity. In this paper, we demonstrate that a temperature compensation mechanism can be devised to solve this problem. We have designed and fabricated a chip implementing a two-layer analog neural network trained to classify low-resolution images of handwritten digits with a low-cost single-poly complementary metal-oxide-semiconductor (CMOS) process, using unconventional analog NVMs for weight storage. We demonstrate a temperature-resilient analog neuromorphic chip for image recognition operating between 10$^{\circ}$C and 60$^{\circ}$C without loss of classification accuracy, within 2\% of the corresponding software-based neural network in the whole temperature range.
\end{abstract}


\section{Introduction}

\IEEEPARstart{R}{eal-time} critical applications require low-latency inference, which can be obtained if artificial neural networks (ANNs) are directly embedded in sensors and in edge devices, with limited compute and energy resources. Inference is computationally demanding and inefficient when performed with classical von Neumann architectures, where processing is separated from memory \cite{XuX2018, Bian2021}. Analog neuromorphic chips based on analog in-memory computing (AIMC) architectures have the potential to overcome the bottlenecks of traditional approaches \cite{Berggren2021, Christensen2022,Zhang2023}, through the maximally parallel one-shot operation of vector-matrix multipliers (VMMs): designers are able to intrinsically embody computing primitives in device physics and in circuit topology, and hence obtain at the same time elegant, compact, and massively parallel computing elements \cite{Zhang2020, Markovic2020, WeiMed2023}. 

Analog neuromorphic chips  use transistors and locally instantiated non-volatile memories operated in sub-threshold conditions \cite{Sun2023} to reach ultra-low power consumption and a large dynamic range \cite{Merrikh-Bayat18, Danial2019, Paliy2020, WangTechnion2022, Rizzo2022_TDVMM}. They can exploit the fact that ANNs can achieve high inference accuracy even under reduced precision \cite{W_Haensch_IEEEProc18} to effectively address the susceptibility of analog computing to noise and non-linearity. This enables analog neuromorphic chips to achieve energy efficiency improvements of up to two orders of magnitude compared to digital accelerators 
\cite{Mehonic2022, Schuman2022, Tye2023, W_Haensch_IEEEProc18}.

Several NVM technologies have recently been considered for AIMC implementations \cite{WeiMed2023, MANNOCCI_APLML_2023}. 
Some of them are based on emerging materials: magnetic tunnel junctions \cite{PatilISCAS2019, Rzeszut_ScRep2022}, ferroelectric devices \cite{Kim2022, Covi_2022, LinEDL23, ThomannIRPS23}, and synaptic memory transistors based on transition-metal dichalcogenides (e.g. MoS$_2$ \cite{Marega2022, Marega2023, Du2017, Farronato2023}). Given that device yield is still limited and integration with CMOS is challenging for some materials, in these cases the full-network operation is implemented only through software-level or mixed-hardware approaches.
On the other hand, multicore accelerators integrating in the same chip both the memristor-based NVM weights (resistive random access memory, RRAM \cite{Huo2022,Wan2022_NeuRRAM,Sun2020,Wang2023} or phase change memory, PCM \cite{KhaddamVLSI2021, LeGalloIBM2023}) along with the other processing circuits, have been realized taking advantage of CMOS compatibility. However, since these types of memory are inherently bistable,  multiple cells are used to store each individual weight to achieve sufficient analog resolution, such as two RRAM cells \cite{ Wan2022_NeuRRAM} or four PCM cells \cite{LeGalloIBM2023}. A sophisticated programming scheme has recently been proposed to improve the programming resolution and reproducibility of RRAMs \cite{RaoMIT2023}, but memory retention is limited by conductance relaxation \cite{Huo2022}. 

Well-established CMOS NVM technologies, consisting of commercial NVM arrays of double-poly NOR flash \cite{Merrikh-Bayat18} or single-poly Y-Flash \cite{yflpat}, have also been proposed for AIMC applications, and are at the basis of larger scale chips by startups in the field \cite{Fick2022}.
The nonlinear current-voltage behavior of such memristive cells is a known challenge: as an example, only binary inputs are accepted by the 12×8 AIMC Y-Flash crossbar array \cite{Danial2019, WangTechnion2022} to avoid introducing distortion in the multiplication operations. 

As is very typical of subthreshold analog circuits, sensitivity of AIMC circuits to temperature is a big challenge: their behavior as a function of temperature is normally reported only for some memory retention tests, while no experimental data have been shown on inference accuracy degradation of ANNs with temperature variation.
As for AICM based on CMOS flash, only few temperature compensation approaches have been investigated through circuit simulations or partial multiplier implementations \cite{Huang10, Guo17, Dilello17, Wei17}, and  similar techniques have been proposed for mixed-signal in memory computing based on SRAM digital weights \cite{Monga23, Seo24, Singh24}.

\begin{figure*}[b!] 
\centering
\includegraphics[width=0.95\linewidth]{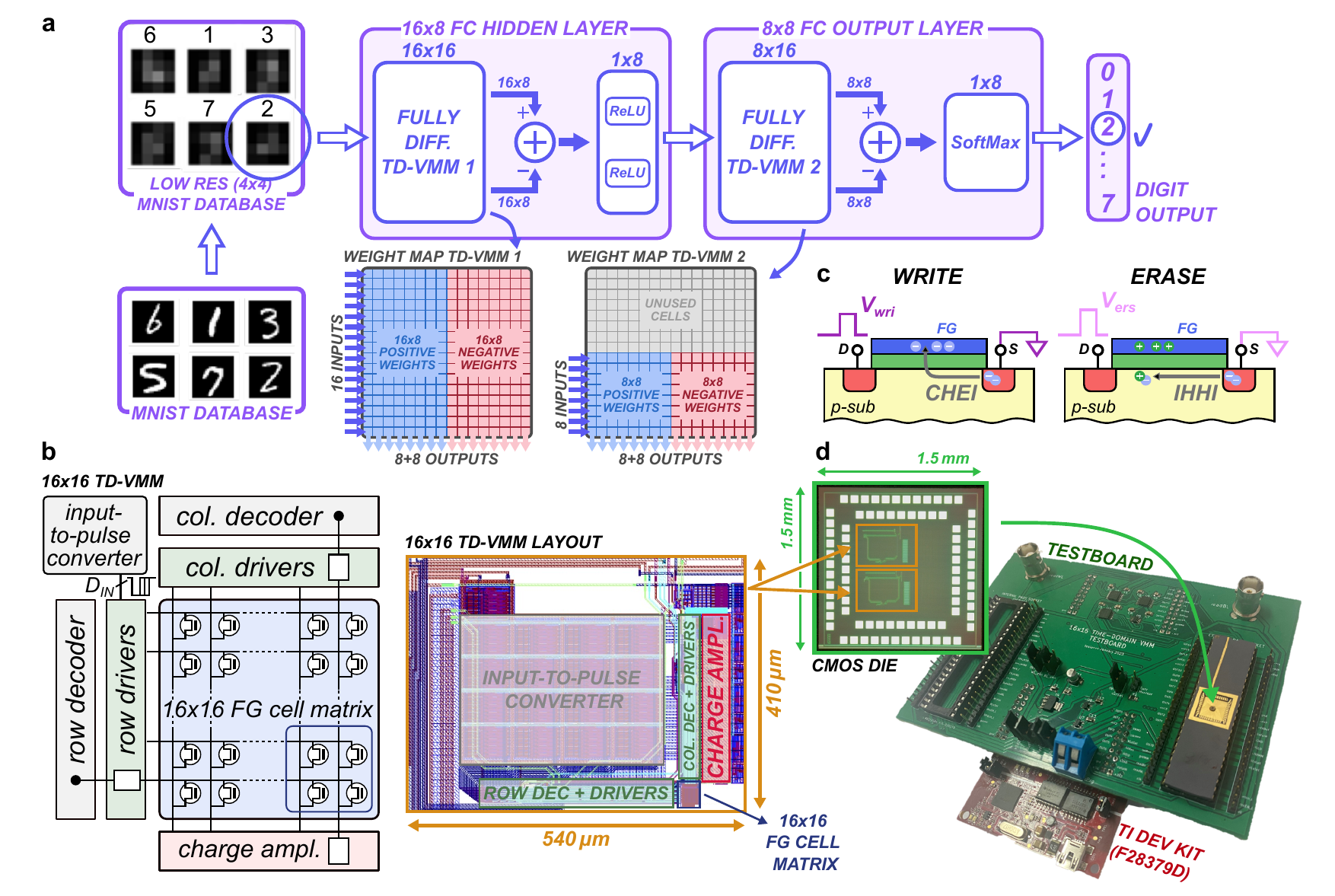}
\caption{Multi-layer fabricated neural network. (a) Architecture of the multi-layer neural network performing image classification on a low-resolution ($4\times4$) version of MNIST database, with detail of the mapped weights on the two TD-VMMs used in the hidden and output fully-connected layers; (b) architecture and layout of the designed analog $16\times16$ time-domain VMM, with details of row and column decoders and drivers, input-to-pulse converter, charge amplifiers and NVM FG cell matrix; (c) sketch of the physical phenomena involved when writing (hot electron injection) or erasing (impact-ionisation hot hole injection) a 1T-FG cell; (d) the fabricated chip was packaged and tested using a custom PCB and commercial development kit.} 
\label{fig1}
\end{figure*}

We have devised a technique to adapt the driving signals of the neuromorphic chip in response to temperature variation in order to compensate for the temperature sensitivity of the analog non volatile memories.
This enables us to demonstrate a temperature-resilient analog CMOS multi-layer ANN, with multi-bit equivalent precision for inputs, weights and outputs of each layer of neurons.

The ANN analog chip, described in Section II, is fabricated with a single-poly 180~nm CMOS process, embedding two $16\times16$ analog time-domain VMMs (TD-VMMs), with the analog weights stored in two-terminal single-transistor floating-gate (1T-FG) memory cells \cite{Rizzo2022_TDVMM}. 
The TD-VMM consists of a $16\times16$ weight crossbar array which is realized with 256 densely placed 1T-FG cells, each obtained using 3.3~V n-type minimum-size transistors with a 7~nm-thick SiO$_2$ gate oxide and a single-poly gate that is left floating (only 1.72~$\rm{\mu m^2}$ per each 1T-FG cell).
We use the 1T-FG cells as programmable current sources: the associated weight is the current flowing through the 1T-FG cell when it is active, i.e., when a constant voltage amplitude $V_{\rm READ}$ is applied between drain and source. In this way the nonlinear current-voltage characteristic of 1T-FG cells does not affect multiplication accuracy, and therefore we experimentally achieve an equivalent number of bits (ENOB) of up to 4.7 bits for the TD-VMM weights and up to 5.7 bits for the TD-VMM outputs.

The temperature compensation method for the VMM and for the  weights stored in the analog non-volatile memory array is described in Section III.
We analyze the intrinsic temperature dependence of the weight stored in the memory array and we devise an adjustment technique for the driving voltages of the VMM that makes the weights almost independent of temperature in a broad temperature range.

As a proof of concept, and to test the inference accuracy on a known benchmark, we have trained the two-layer ANN for the classification of a low-resolution ($4\times4$ pixels) version of the handwritten digit MNIST database \cite{MNIST}, reaching a classification accuracy of 83.1\% at the temperature of $30^{\circ}\rm{C}$ (results are shown in Section IV). We have rigorously characterized the chip to assess its temperature sensitivity and the time retention of the analog NVM. We verify that a classification accuracy higher than 81.5\% can be achieved in a $10\div60^{\circ}\rm{C}$ temperature range (81.56 \% at 10$^{\circ}\rm{C}$, 83.03\% at 60$^{\circ}\rm{C}$), thanks to the adaptive compensation of $V_{\rm READ}$ in response to temperature variation, and for 7 days at 20$^{\circ}\rm{C}$, or for more time if a memory refresh is performed on a weekly basis.

\section{Analog Neural Network Chip}


\begin{figure*}[b!] 
\includegraphics[width=\linewidth]{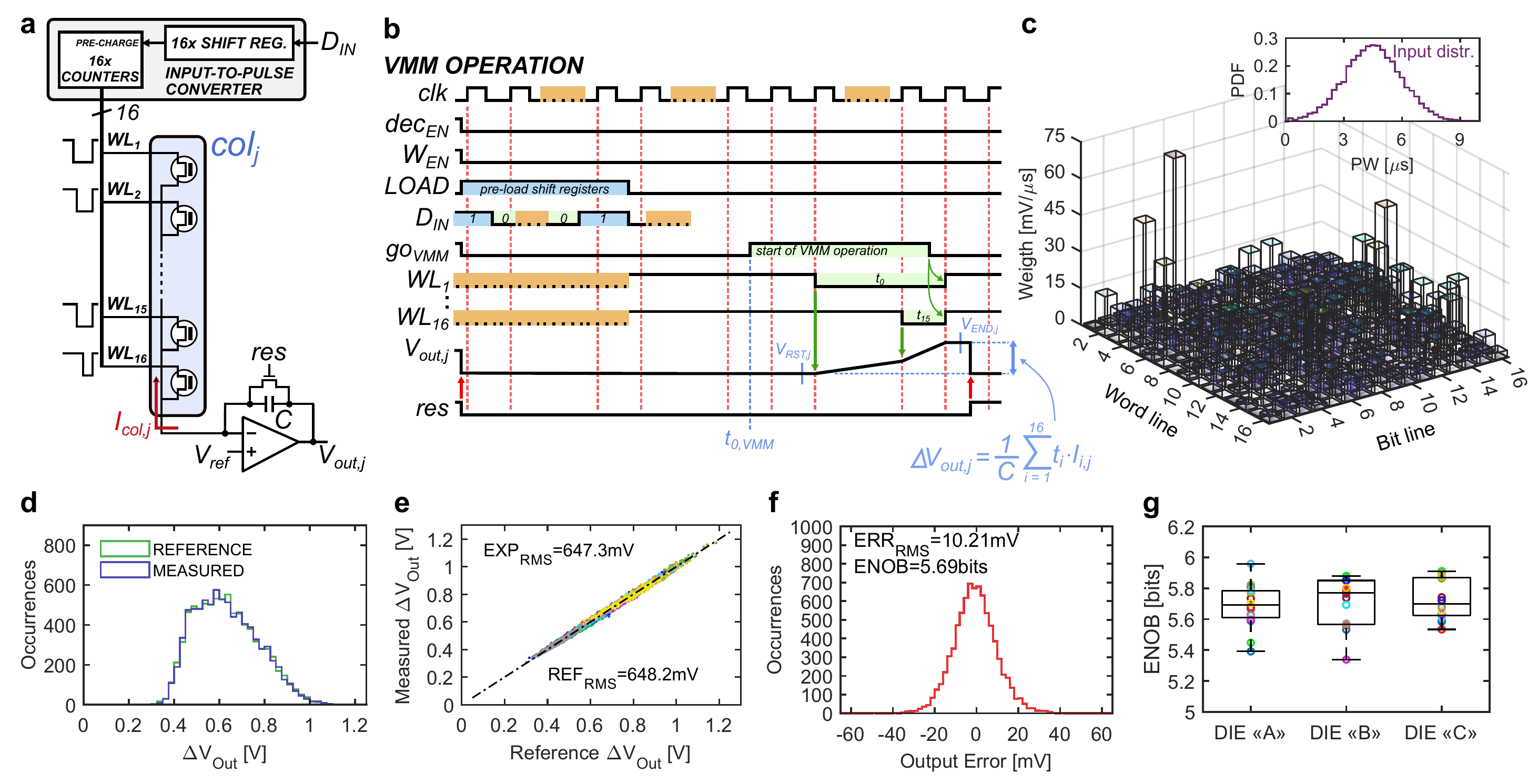}
\caption{TD-VMM operation and precision characterisation: (a) Detail of the MAC operations performed by a column (BL) of the TD-VMM; (b) timing diagram of the VMM operation; (c) spatial map of the native weights of a typical $16\times16$ crossbar array (random pulse-width input vector distribution shown in the inset); (d) histograms of the measured and theoretical VMM outputs for a given set of random input data, with (e) corresponding scatter plot of measured vs. theoretical outputs (f) and extracted output errors ($\text{ERR}= \Delta V_{\rm OUT,EXP}-\Delta V_{\rm OUT,REF}$); (g) ENOB extracted for each BL column of three different dies.} 
\label{fig2}
\end{figure*}

\subsection{Analog neural network architecture}

The architecture and the concept of operation of the analog neural network are shown in Fig.~\ref{fig1}a: the inputs are low-resolution images of handwritten digits from `1' to `8' ($4\times4$ pixels, 32-tone grayscale representation), which are a reduced version of the MNIST dataset images. 
The input image is passed as an 80-bit [($4\times4$)~pixels~$\times$~5~bits] digital stream to the analog chip, where it is converted into 16 voltage pulses (one per pixel) of fixed amplitude $V_{\rm READ}$ and width proportional to the gray level, which represent the input vector of the first fully-connected $16\times8$ hidden neuron layer, consisting of an analog TD-VMM followed by a Rectified Linear Unit activation function acting on the 8 sampled  outputs; after that, the resulting data are processed by a fully-connected $8\times8$ output neuron layer, again consisting of a second analog TD-VMM followed by a softmax activation function that performs the final 8-digit classification. 

To address the need of both positive and negative weights, Fig.~\ref{fig1}a illustrates the mapping schemes adopted for both hidden- and output-layer weights. To map the positive and negative weights of the $16\times8$ reference (target) hidden layer into the corresponding hardware TD-VMM, we exploit the full $16\times16$ matrix of the first TD-VMM to write positive weights on its left half (with a value of `0' in correspondence to the location of negative weights) and the absolute value of negative weights on its right half (with a value of `0' in correspondence to positive weights). Therefore, we are able to perform the $16\times8$ operations -- with signed number representations -- by subtracting the 8 right-side outputs from the left-side ones. Similarly, the reference output-layer positive and negative weights are mapped into an 8×16 matrix within the second hardware TD-VMM, to address the related signed-representation $8\times8$ VMM operation. For the sake of clarity, from now on we distinguish between unsigned (1T-$W$) and signed (2T-$W$) analog weights, as obtained by the difference between two unsigned 1T-$W$ cells implementing the same weight. 

Fig.~\ref{fig1}b focuses on the top-level architecture and layout of the analog TD-VMM: its core is the $16\times16$ memory matrix realized with single-poly 1T-FG cells, together with an array of 16 charge amplifiers -- one for each column -- where the compute-in-memory operation takes place \cite{W_Haensch_IEEEProc18}. In addition, an array of digital input-to-pulse converters is used to feed the input pulse data, and row/column decoders and drivers enable the implementation and selection of the proper operation mode (comprising also the programming and reading of the NVM matrix). Each cell can be selected via row and column decoders, whereas row and column drivers are exploited to properly bias selected and non-selected cells.

The analog weights are programmed (Fig.~\ref{fig1}c) using a series of voltage pulses of different amplitudes on the drain, with the source grounded, using a program and verify scheme \cite{Rizzo2022_TDVMM}. Some more details are discussed in {\it Appendix A}.
Fig.~\ref{fig1}d shows the experimental chip that includes the two TD-VMMs, with dual bonding option for debug purposes. In addition, a test setup, composed of a custom printed circuit board with die socket and of a commercial evaluation board (TI LAUNCHXL-F28379D), was implemented to run the characterization and testing of the multi-layer ANN chip.

We limit our analysis to the proof-of-concept network of Fig.~\ref{fig1}a due to the available hardware, as our demonstrator chip contains two $16\times16$ VMMs. Let us stress that the proposed temperature compensation mechanism is based on a general concept that can be effective also on larger networks. Future works will address the design and fabrication of larger networks, enabling  classifiers for high-resolution datasets and addressing practical scaling up issues and half-selection issues (discussed more in detail in {\it Appendix A}).

\begin{figure*}[b!] 
\centering
\includegraphics[width=0.95\linewidth]{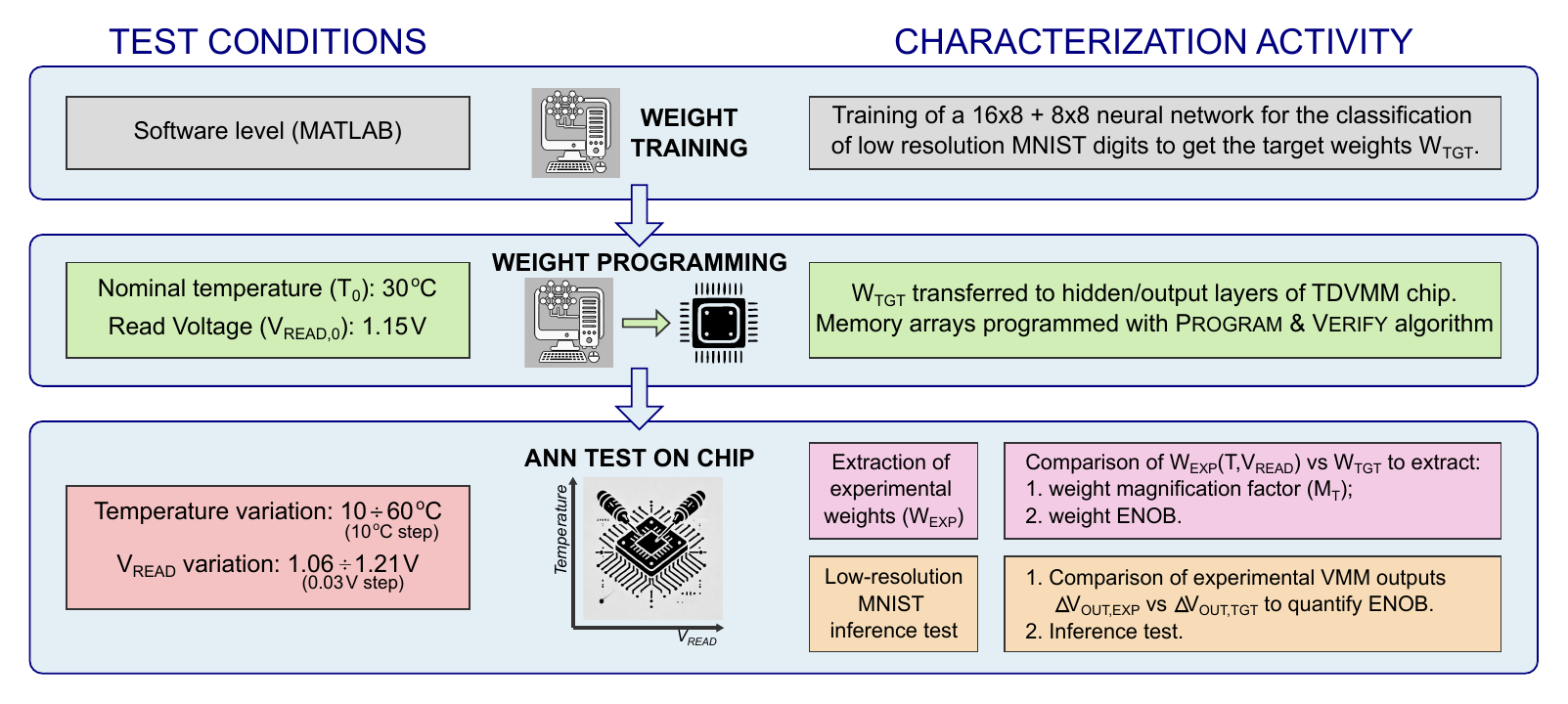}
\caption{Flowchart of the tools and techniques used in the present work.}
\label{fig3new}
\end{figure*}

\subsection{Time-Domain Vector-Matrix Multiplier}

To illustrate the operation of the proposed TD-VMM, Fig.~\ref{fig2}a shows the details of a VMM column implementation.
The inputs are provided as voltage pulses at each row -- word-line (WL) -- connected to the 1T-FG cell source terminals. These pulses  have an amplitude that ensures a constant drain-to-source voltage $V_{\rm READ}$ is applied to the corresponding 1T-FG cell, and a pulse width corresponding to the analog input data. As a result, the drain currents of the 1T-FG cells have a rectangular-shape waveform: for each transistor cell, the on-current level is the analog weight, whereas the integral of the corresponding current over time (i.e., the total charge) represents the  physical product between the pulse width (proportional to the pixel gray level) and the on-current (proportional to the weight). The drain terminals of all 1T-FG cells in the same column are connected through the column driver to the corresponding bit-line (BL) and to the input of the related charge amplifier: the sum of the charges flowing through the cells in the same column $j$ (i.e. the sum of the products of analog inputs by analog weights) is conveyed at the inverting input of the charge amplifier, and is converted into an output voltage ($V_{{\rm out},j}$). 
With reference to Fig.~\ref{fig2}a, showing the details of a VMM column implementation during VMM operation, the decoders are deactivated, since all the $16\times16$ matrix cells operate in parallel; the drivers make sure that each row -- word-line (WL), connected to the 1T-FG cell source terminals -- sees its relative input time pulse, which activates with a fixed drain-to-source voltage the corresponding row of 1T-FG cells for the time of its duration. The timing diagram of the VMM operation is depicted in Fig.~\ref{fig2}b and described in detail in {\it Appendix B}.

\subsection{Temperature Characterization Methodology}

Fig.~\ref{fig3new} reports a flowchart highlighting tools and techniques used in the present work: we have trained a two layer neural network in Matlab and the trained weights have been programmed into the chip hidden- and output-layer at  nominal temperature $T_0$ = 30$^{\circ}$C and $V_{\rm READ} =  1.15 ~\rm{V}$. 
As the weights vary exponentially when changing the temperature, the temperature compensation mechanism exploits the variation of the $V_{\rm READ}$ to adjust the weight values. Therefore, we have evaluated the experimental weights programmed in the previous step by varying the temperature in the $10 \div 60 ^{\circ}$C range and the $V_{\rm READ}$ in the $1.06\div1.21 ~\rm{V}$ range. In this way, by plotting extracted weights ($W_{\rm EXP}$) versus target ones ($W_{\rm TGT}$), we have been able to quantify the magnification $M_T$ (first order effect) and the $\rm{ENOB}$, using the definition inherited from the analog-digital converter characterization protocol \cite{SINAD}: $(\rm{SNDR}_{dB}-1.76)/6.02$ , where $\rm{SNDR}$ (signal to noise and distortion ratio) is here calculated as the ratio of the $W_{\rm TGT}$ RMS value to the $\Delta W = W_{\rm EXP}/M_T-W_{\rm TGT}$ RMS error, which is caused by both noise and nonlinearity. 
For the same temperature and  $V_{\rm READ}$ conditions, the implemented neural network has been fully characterized by processing all the low-resolution MNIST images of the test dataset, and the output results have been extracted and compared with the correct results.

\section{Temperature compensation of analog weights and of  VMM operation}

\begin{figure*}[b!]%
\centering
\includegraphics[width=\textwidth]{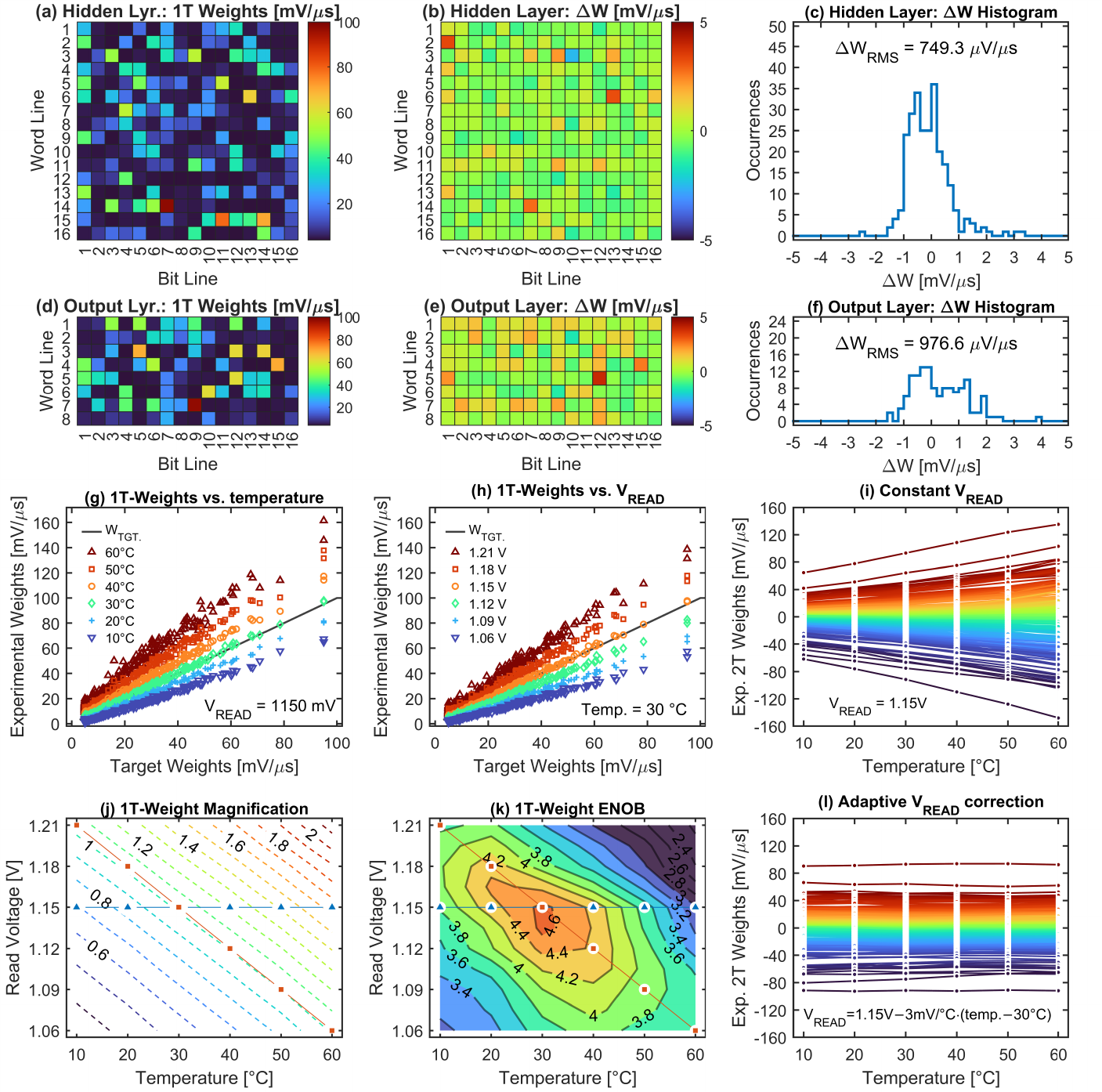}
\caption{Programmed Weights. Hidden Layer: spatial maps of the \textbf{(a)} experimental 1T-weights and \textbf{(b)} corresponding $\Delta W = W_{\rm EXP} - W_{\rm TGT}$ and \textbf{(c)} related error histogram. Output Layer: spatial maps of the \textbf{(d)} experimental 1T-weights and \textbf{(e)} corresponding $\Delta W$ and \textbf{(f)} related error histogram. Scatter plots of $1T-W_{\rm EXP}$ (hidden+output layers) with respect to $1T-W_{\rm TGT}$ as a function of \textbf{(g)} temperature  and of \textbf{(h)} $V_{\rm READ}$. \textbf{(j)} Weight magnification and \textbf{(k)} ENOB as a function of temperature and $V_{\rm READ}$. 2T-weight dependence of the temperature under \textbf{(i)} constant $V_{\rm READ}$ and \textbf{(l)} $V_{\rm READ}$ correction  conditions.}
\label{fig3}
\end{figure*}

Due to the analog nature of the neuromorphic chip, CMOS temperature dependent parameters could undermine the inference accuracy of the neural network,  when the actual temperature $T$ deviates from the nominal temperature $T_0$, i.e. the temperature of the chip when the analog weights have been programmed with the program and verify scheme.

Although FG memory cells operating in the sub-threshold regime are highly sensitive to temperature, only few papers investigate compensation approaches, typically through circuit simulations or partial implementations: in \cite{Huang10}, a translinear loop-based FG current reference circuit was proposed, achieving linear FG charge-to-current proportionality while reducing temperature sensitivity; however, each cell is large (consisting of one FG cell, four transistors, and one resistor) and cannot be considered a viable option for large density VMMs. Refs. \cite{Guo17} and \cite{Dilello17} investigate current mirror FG cell topologies, with weights embedded in the current magnification factors with respect to the input current of a reference FG cell. Since the control gate voltage is not fixed but it is generated by the input cell, it is dynamically adjusted in reaction to temperature variations. However, besides the special case of unitary current gain, significant sensitivity remains for large or small weights. Ref. \cite{Wei17} presents an oversimplified analysis of an AIMC computing chip, using simulations on standard 180 nm CMOS transistors to emulate FG cells; the proposed VMM is implemented with transistors operating in the linear region, mapping weights onto the threshold voltage difference of two coupled transistors (it is not clear how the threshold voltage $V_{th}$ is programmed without an FG) and inputs to the drain-source voltage. By making the difference of the output currents to get the output result, they mitigate $V_{th}$ variations, leaving only mobility dependence on temperature. Other temperature compensation strategies have been proposed for mixed-signal in memory computing based on SRAM digital weights \cite{Monga23, Seo24, Singh24}. 
Regarding alternative weight memory representation methods, different memory cell technologies exhibit distinct temperature-dependent behaviors due to the varying physical mechanisms governing conduction, therefore require ad hoc compensation techniques.

Fig.~\ref{fig3} analyzes how accurately the target weights $W_{\rm TGT}$ are mapped into the two TD-VMM memory cores as experimental analog weights $W_{\rm EXP}$, with emphasis on temperature behavior; a voltage amplitude adaptive compensation is then proposed to improve weight accuracy.

At nominal temperature $T_0$, analog weight programming grants good accuracy, as confirmed by the spatial maps and histograms of programmed weights and corresponding errors (defined as the difference between $W_{\rm EXP}$ and $W_{\rm TGT}$), presented in Fig.~\ref{fig3}a-c for the hidden layer ($16\times16$) and in Fig.~\ref{fig3}d-f for the output layer ($8\times16$): the error RMS values remain lower than $750~\rm{\mu V/\mu s}$ for the hidden layer and than $980~\rm{\mu V/\mu s}$ for the output layer.

However, as the temperature changes, the experimental weights change drastically, since 1T-FG cells operate in sub-threshold, where the current varies exponentially with the temperature due to the linear decrease of threshold voltage $V_{th}$ with temperature of about $-1~\rm{mV/^{\circ}C}$ \cite{Tzou1985}. The scatter plot of Fig.~\ref{fig3}g highlights this dependence, showing how the weight error can exceed 75\% as $T$ rises by only 30$^{\circ}$C above $T_0$. The impact of temperature can be attributed to three main effects: weight magnification, distribution bending, and spreading.

We exploit an adaptive variation with temperature of the voltage amplitude $V_{\rm READ}$ to compensate for such degradation: according to Fig.~\ref{fig3}h, obtained for $T=30^{\circ}$C, the effect of the voltage amplitude variation on the weights is similar but opposite with respect to the temperature. This is essentially due to sub-threshold biasing of the 1T-FG transistors, where the current is proportional to $\exp[(V_{GS-{\rm READ}}-V_{th})/(m k_B T)]$, where $V_{GS-{\rm READ}}$ is the partition of $V_{\rm READ}$ applied between gate and source through capacitive coupling, $k_B$ is Boltmann's constant and $m$ is the MOSFET body effect coefficient.


The benefits of an adaptive voltage amplitude are well visible taking Fig.~\ref{fig3}i as a starting point, which shows the $W_{\rm EXP}$ strong temperature dependence at constant $V_{\rm READ}$, for both hidden and output layers. We were able to quantify weight magnification and ENOB degradation at varying temperature and $V_{\rm READ}$ in Fig.~\ref{fig3}j and k, respectively: a temperature-insensitive unitary magnification factor, as well as a partial weight-ENOB compensation, can be obtained if the $V_{\rm READ}$ is linearly scaled with temperature, with a ${\Delta V_{\rm READ} / \Delta T}$ correction of $-3~\rm{mV/^{\circ}C}$, as indicated by the red lines in Fig.~\ref{fig3}j,k. After correction, the new weight curves are reported in Fig.~\ref{fig3}l, showing a clear and beneficial impact on the weight resilience to temperature.

The impact of temperature of weight transformation is composed of linear and non-linear (weight-dependent) factors. For a given weight value, the almost linear magnification observed in Fig.~\ref{fig3}i (and quantified in Fig.~\ref{fig3}j) represents the dominant, first-order effect. In Fig.~\ref{fig3}g, as second order effects, a slight nonlinear distribution bending is observed, as the weight at a temperature different from $T_0$ gets distorted through a nonlinear factor ($W_{[T]} \propto (W_{[T0]})$\textsuperscript{$T_0$/$T$}). In addition, we have observed a weight spread in temperature: if we program two cells to the same $W_{[T0]}$ value, their $W_{[T]}$ values can differ due to a mismatch in the temperature response (this is a non-deterministic behavior, thus it cannot be compensated as it is unpredictable). 
To quantify linear magnification on one side, and non-linear magnification and random spread on the other side, we have extracted the magnification factor and weight ENOB in Fig.~\ref{fig3}j and k, respectively. Note that the weight ENOB is calculated after performing a linear fit of the experimental weights to the target values, in order to isolate only second-order effects. The observed weight ENOB degradation can be thus ascribed to just second-order effects. Thus, although the slight curvature bending is not visually prominent in the scatter plots in Fig.~\ref{fig3}g, it is quantitatively captured by the ENOB plot in Fig.~\ref{fig3}k.

Our compensation technique is thus effective in tackling both first and second order effects: it maintains an optimized weight scaling across a broad temperature range (Fig.~\ref{fig3}j) while partially correcting the distortions introduced by second-order effects (Fig.~\ref{fig3}k).
Further details on these aspects are described in {\it Appendix~C}.

\section{Experimental Demonstration of Temperature-resilient Classification Accuracy}

\begin{figure*}[t!]%
\centering
\includegraphics[width=\textwidth]{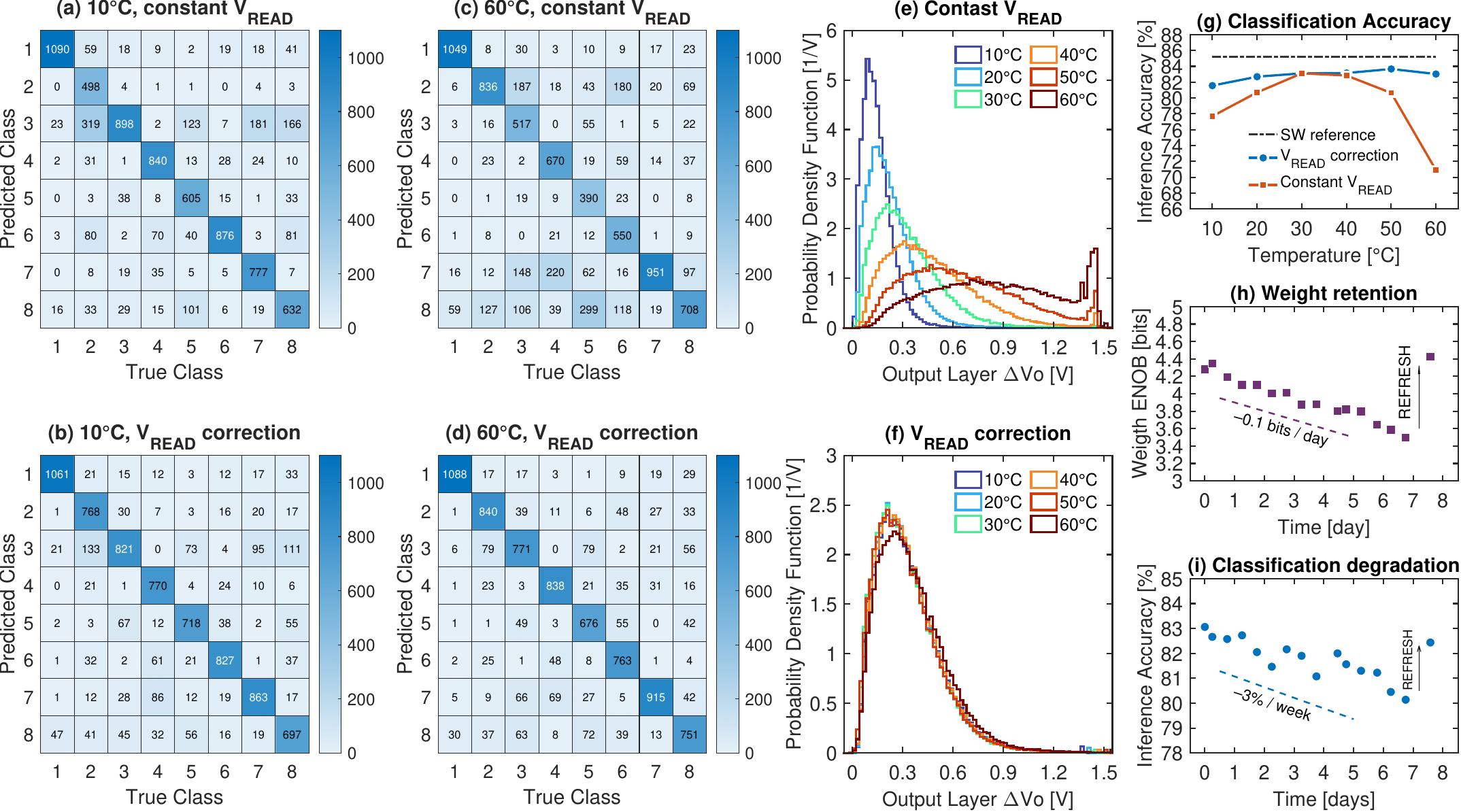}
\caption{\textbf{Low-resolution digit classification.} Confusion matrices extracted under constant $V_{\rm READ}$ and adaptive $V_{\rm{READ}}$ tests at 10$^{\circ}$C (\textbf{(a)} and \textbf{(b)}, respectively) and 60$^{\circ}$C (\textbf{(c)} and \textbf{(d)}, respectively).  Corresponding normalized histograms (in probability distribution function form) of the output-layer TD-VMM outputs under \textbf{(e)} constant $V_{\rm READ}$  and \textbf{(f)} $V_{\rm READ}$ correction tests. \textbf{(g)} Classification accuracy against temperature. Degradation over time of the \textbf{(h)} weight ENOB  and of the \textbf{(i)} network classification accuracy.} .
\label{fig4}
\end{figure*}

The low-resolution (4$\times$4) MNIST test set presented in Fig.~\ref{fig1}, with digits from `1' to `8', has been used as a testbench for the characterization of the full neural network chip. The trained ANN parameters have been mapped into two TD-VMM cores within our CMOS chip, by programming the hidden-layer and output-layer weights (Fig.~\ref{fig3}a and d, respectively) at $T=30^{\circ}$C and $V_{\rm READ} = 1.15~\rm{V}$. The programming operations were performed by targeting a weight ENOB of 4.5 bits (see Fig.~\ref{fig3}k) in the nominal conditions.

Two main inference test analyses have been performed, to verify the impact of temperature- and time-dependent degradation of the accuracy performance.
As for the temperature behavior, we have tested the chip inference accuracy in the $10\div60^{\circ}$C range, by comparing two approaches: on one side, we have varied the temperature by keeping all signal scaling factors and $V_{\rm READ}$ constant, to emphasize the impact of the natural temperature-sensitivity of the 1T-FG memory cells storing the weights; on the other side, the same analysis has been repeated by using the adaptive $V_{\rm READ}$ with the linear correction rule (${\Delta V_{\rm READ} / \Delta T}=- 3\rm{mV/^{\circ}C}$). 

As a preliminary remark, one should note that, in principle, a
linear scaling of the weights would not affect the classification
performance in the implemented neural network architecture.
However, when the weights magnification is not accompanied
by a corresponding rescaling of the input data, the integrator
circuits are pushed towards saturation, as limited by the supply
voltage. In addition, a further non-linear effect is introduced
by the fact that weight transformation with temperature is, in fact, not exactly linear (see again Fig. 3k).

Fig.~\ref{fig4}a-d report the confusion matrices of the individual handwritten digit tests at 10$^{\circ}$C (a,b) and 60$^{\circ}$C (c,d), comparing the two cases of constant and corrected $V_{\rm READ}$, where we see a clear reduction of inference errors when using an adaptive voltage amplitude. In the constant $V_{\rm{READ}}$ mode, at 10$^{\circ}$C (Fig.~\ref{fig4}a), the degradation is mainly attributed to images reporting the digit `2', which are often erroneously classified as a `3' by the network (in 319 cases, representing a -4\% of inference accuracy loss). Instead, at 60$^{\circ}$C (Fig.~\ref{fig4}c), the mistakes are distributed within all the digits (e.g. 299 `5' are classified as `8', 220 `4' are classified as `7', and so on). Note that the occurrences of all these typical mistakes are reduced by a good amount when the $V_{\rm{READ}}$ correction method is implemented (Fig.~\ref{fig4}b and d).

For all the investigated temperatures, Fig.~\ref{fig4}e reports the histograms of the output-layer VMM outputs in the fixed-$V_{\rm READ}$ condition, making it clear how the temperature causes a dramatic deformation of the expected results if no countermeasures are taken: low temperature brings the output distribution towards low voltages (blue histogram), thus lowering the signal-to-noise ratio; on the other hand, the upper tail of the distribution at high temperature falls in the saturation operation of the charge amplifiers (brown histogram), thus introducing a high degree of distortion. On the contrary, when the $V_{\rm READ}(T)$ correction is exploited, the histogram of the outputs taken at all the investigated temperatures are well overlapped, as evidenced by Fig.~\ref{fig4}f.



The measured inference accuracy, as obtained by testing 8000 images of the MNIST test-data low-resolution version, is reported in Fig.~\ref{fig4}g, and reaches a typical value of 83\% at nominal conditions (versus an 85\% for a double-precision floating-point software implementation). Without the adaptive $V_{\rm READ}(T)$, the classification accuracy of our chip falls down to 77.7\% at 10$^{\circ}$C and to 70.9\% at 60$^{\circ}$C. However, the adaptive $V_{\rm READ}(T)$ results in a temperature-resilient  accuracy, with all values in the $30^{\circ}$C$\div60^{\circ}$C range slightly higher than 83\% and a minimum of 81.56\% measured at 10$^{\circ}$C.

Finally, the classification accuracy performance of the programmed chip has been measured over time for a week at 20$^{\circ}$C, to investigate the retention of the memory cores storing the analog weights. Fig.~\ref{fig4}h report the ENOB extracted considering both hidden and output weights: an ENOB degradation of approximately $-0.1$~bits/day has been measured, being the cause of the $-3\%$/week classification accuracy degradation shown in Fig.~\ref{fig4}i. The physical nature of the FG memory core allows us to make a memory refresh, which can be exploited to bring the weights back close to the target, leading to a recovered classification capability of the neuromorphic chip. A silicon process with a thicker oxide (e.g. 10~nm) would provide industrial-grade retention.

\section{Conclusion}
Our results show that it is possible to use proper circuit and device design techniques to overcome some of the main weaknesses of analog computing circuits, i.e. sensitivity to non linearity, noise and temperature variations, while still preserving the main strengths of analog computing, i.e. low-power operation and the intrinsic massive parallelism, while providing sufficient precision for achieving high classification accuracy in analog neural network classifiers. Therefore, we believe these results can pave the way towards the exploitation of low-cost CMOS foundry processes to design and manufacture analog neuromorphic chips. In addition, this demonstration and the reliability and accessibility of consolidated CMOS processes are a very good starting point to achieve very-low-energy per inference in large artificial intelligence models implemented with analog neuromorphic chips.



\appendix

\begin{figure*}[b!]%
\centering
\includegraphics[width=0.9\textwidth]{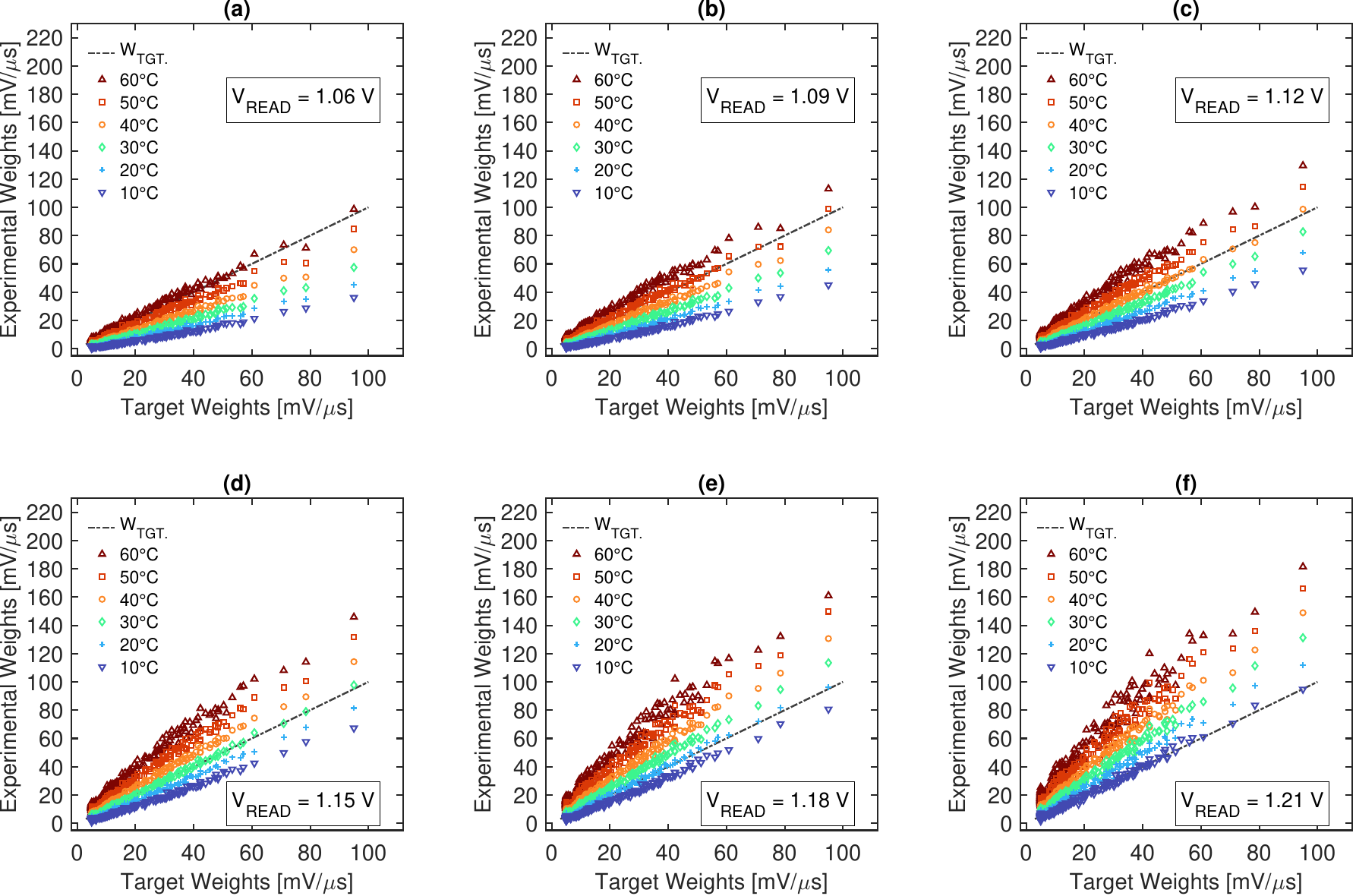}
\caption{\textbf{Hidden Layer: weight scatter plots for different $V_{\rm READ}$ and temperature conditions.} \textbf{(a)}  $1.06~\rm{V}$, \textbf{(b)} $1.09~\rm{V}$, \textbf{(c)} $1.12~\rm{V}$, \textbf{(d)} $1.15~\rm{V}$, \textbf{(e)} $1.18~\rm{V}$, \textbf{(f)} $1.21~\rm{V}$ (temperature from $10^{\circ}$C to $60^{\circ}$C, with $10^{\circ}$C step).}
\label{fig_ex_1}
\end{figure*}

\begin{figure*}[t!]%
\centering

\includegraphics[width=0.9\textwidth]{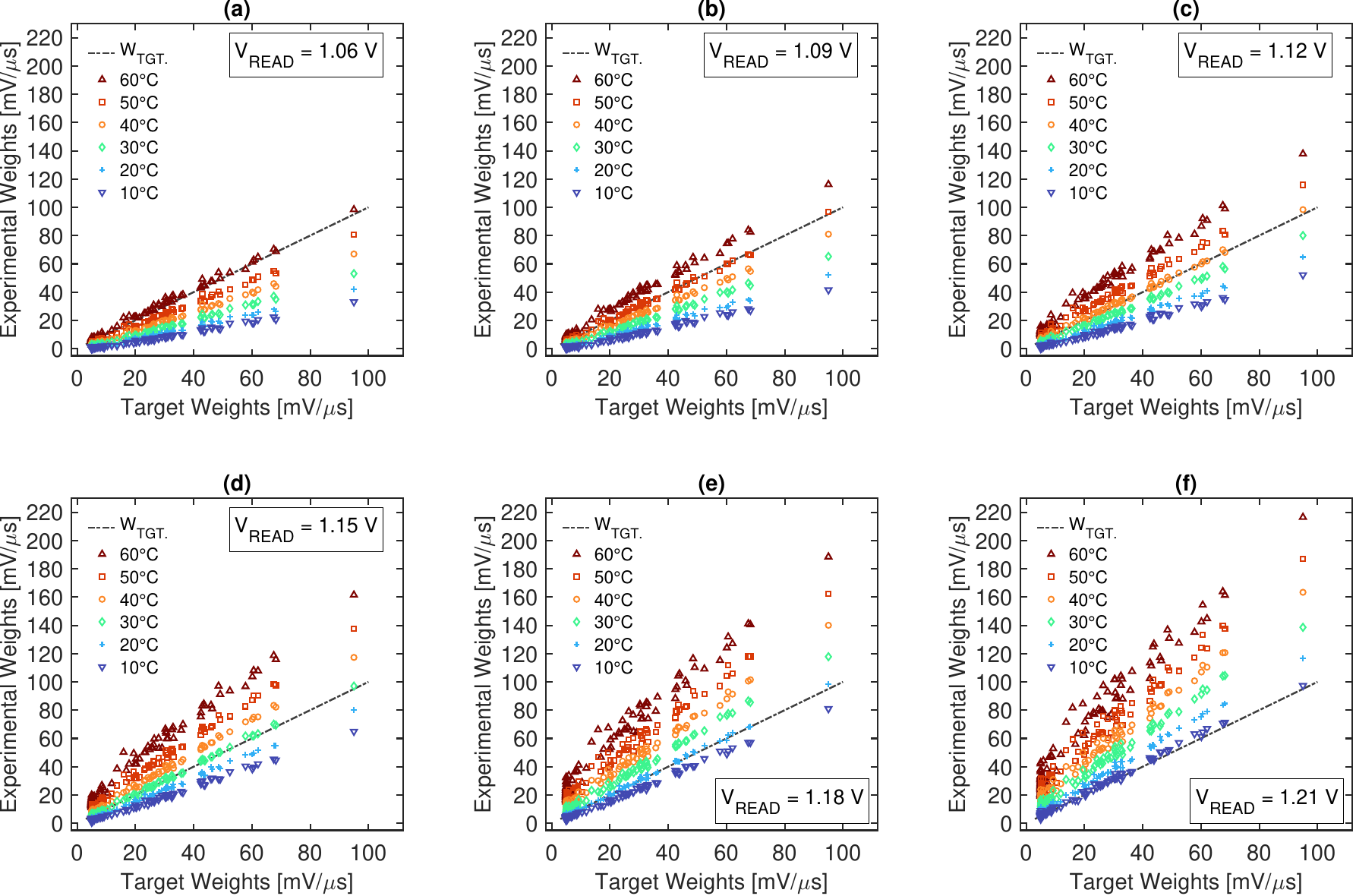}
\caption{\textbf{Output Layer: weight scatter plots for different $V_{\rm READ}$ and temperature conditions.} \textbf{(a)}  $1.06~\rm{V}$, \textbf{(b)} $1.09~\rm{V}$, \textbf{(c)} $1.12~\rm{V}$, \textbf{(d)} $1.15~\rm{V}$, \textbf{(e)} $1.18~\rm{V}$, \textbf{(f)} $1.21~\rm{V}$ (temperature from $10^{\circ}$C to $60^{\circ}$C, with $10^{\circ}$C step).
}
\label{fig_ex_2}

\end{figure*}

\subsection{Analog NVM 1T-FG cell programming and read}

Fig.~\ref{fig1}c illustrates the programming operation for a single 1T-FG cell. A write pulse adds negative charge to the floating polysilicon gate, in order to reduce the cell conductivity and therefore the stored weight: this is obtained by applying a drain voltage $V_{\rm wri}$ between 5.5~V and 6.5~V, causing channel hot electron injection (CHEI) into the polysilicon gate, where electrons are trapped (Fig.~\ref{fig1}c, left). An erase pulse adds positive charge to the floating polysilicon gate, to neutralize the negative charge and to increase cell conductivity (increase the stored weight): a drain voltage $V_{\rm ers}$ higher than 7~V causes impact-ionizing hot hole injection (IHHI) into the floating gate (Fig.~\ref{fig1}c, right). In all cases, the source is at ground\cite{Rizzo2022_TDVMM}.

In order to program the analog weights, we adopt a program and verify scheme, cyclically applying short write/erase pulses and extracting the weights for verification, and then repeating the cycle until the desired current is obtained. 
The weight cells are prone to programming half-selection disturbance due to the crossbar memory architecture, which does not include selectors to isolate unaddressed cells. While this issue is somehow tolerable in relatively small arrays, and can be corrected with a successive-refinement program and verify iterative algorithm, it is likely to become critical in larger crossbar array implementations. In such cases, we believe that an isolation mechanism, such as a selector for each weight cell, would be necessary to prevent cross-coupling during programming operations. 
In this case, we have exploited a partial countermeasure during weight extraction, which is carried out this way: i) a set of random pulse-width input vectors ($PW_{IN}$) are sent to the chip to perform the VMM operations; ii) based on the measured results (analog $\Delta V_{\rm OUT,EXP}$), the weight matrix $W$ is extracted by finding the argument $W_{\rm EXP}$ which minimizes the function $RMS(\Delta V_{\rm OUT,EXP} - PW_{IN} \times W_{\rm EXP})$. Interestingly, since these $PW_{IN}$ input vectors are drawn from a distribution with a  shape similar to that used during the actual inference test (i.e. we used images from the training dataset), this approach helps to compensate for some of the systematic offsets and non-idealities. It can thus be regarded as a form of hardware-aware weight transfer.


\subsection{Time Domain Vector-Matrix Multiplier Operation}

The TD-VMM cores can operate in three different modes: ``VMM'' for normal VMM compute-in-memory operation; ``PROGRAM'' and ``READ'' for memory multi-level programming and reading.
For a closer look at the ``VMM'' mode, we can refer to Fig.~\ref{fig2}a, showing the details of a VMM column implementation. In this operation mode, the decoders are deactivated, since all the $16\times16$ matrix cells operate in parallel; the drivers make sure that each row -- word-line (WL), connected to the 1T-FG cell source terminals -- sees its relative input time pulse, which activates with a fixed drain-to-source voltage the corresponding row of 1T-FG cells for the time of its duration.
Details on the timing of the VMM operation are given in Fig.~\ref{fig2}b, hereafter further discussed.

\textbf{``VMM'' mode - Input timing:} 
The 16 input pulse-width signals are generated by an array of digital-to-pulse converters. The input data are loaded in 16 8-bit registers (with signal \textit{LOAD} active), and are then transferred to 16 8-bit counters. The MSB of each register and counter is always pre-loaded with `1', while the other 7 bits (which we name \textit{D}$_{IN,i}$, with `i' from 1 to 16) contain the desired i-th pulse duration value in terms of clock periods, $T_{\rm CLK} = 250~\rm{ns}$ (minimum limited by the evaluation board microcontroller, as a standard GPIO pin is used to provide the clock to the analog chip). When the \textit{LOAD} signal is deactivated and the VMM operation starts (low-to-high transition of the $go_{\rm VMM}$ signal, at time $t_{\rm 0,VMM}$) the counters start counting and each corresponding \textit{MSB}$_{i}$ goes low after $(128-\textit{D}_{IN,i}) \times T_{CLK}$. The 16 \textit{MSB}$_{i}$ lines are buffered and used to implement the 16 pulse-width WLs, as they stay low for a time equal to \textit{D}$_{IN,i} \times T_{\rm CLK}$, being reset to `1' all together at the end of 128 $T_{\rm CLK}s$ after $t_{\rm 0,VMM}$.

\textbf{``VMM'' mode - Output timing:} The corresponding output voltage waveform of a column integrator is also shown. The integrator is reset at the beginning of each VMM operation (with \textit{res} signal), and its output voltage has a piece-wise-linear waveform with the slope increasing at the beginning of each WL pulse.
The $V_{{\rm OUT},j}$ of each column is sampled before the integration  (after reset) -- $V_{{\rm RST},j}$ -- and at the end of the operation -- $V_{{\rm END},j}$ -- , and the $j$-th output result of the VMM is given by the voltage difference $\Delta V_{{\rm OUT},j} = V_{{\rm END},j} - V_{{\rm RST},j}$.

\textbf{``PROGRAM'' mode:} Fig.~\ref{fig1}c displays the 1T-FG single cell architecture highlighting the physical phenomena which enable program and erase operations.
In ``PROGRAM'' mode, the BL is forced to ground, whereas a programming pulse is provided to the WL (with amplitude $V_{\rm prg}$). Non-selected WLs and BLs are kept in high impedance to limit half-selection issues. The programming voltage pulse $V_{\rm prg}$ can activate different injection phenomena depending on its intensity, thus writing or erasing the cell. For a write operation, a $V_{\rm wri}$ voltage as high as 5.5~V to 6.5~V and typical pulse width of 20~ms causes the tunneling of hot electrons from the channel to the gate, where they remain trapped, leading to an increase of the threshold voltage $V_{th}$ of the cell. For even higher $V_{\rm prg}$ values, erasing occurs, since the accelerated electrons generate electron-hole pairs in the channel and the high $V_{DS}$ favors hole injection through the gate oxide into the floating gate, resulting in a $V_{th}$ reduction. To get an effective erase we have used 9~V $V_{\rm ers}$ pulses of 100~$\rm{\mu s}$. 
1T-FG cell program is a demanding operation from both time and energy point of view. Total programming time, $T_{\rm Full-Prog}$: tens  of  iterations $\times$ number of weights $\times$ $t_{\rm prog}$, with typical $t_{\rm prog}$ of $100~\rm{\mu s}$ (reset) up to $20~\rm{ms}$ (set); total programming energy, $E_{\rm Full-Prog}$: $T_{\rm Full-Prog}$ $\times$ $V_{\rm prog}$ $\times$ $I_{\rm prog}$, with typical  $V_{\rm prog}$ and $I_{\rm prog}$ of 6~V (set) to 9~V (reset) and $\sim 1~\rm{mA}$, respectively.

\textbf{``READ'' mode:} the ``READ'' mode enables the physical access to the terminals of each single cell, as the selected WL and BL are connected to the external circuitry via the read$_{WL}$ and read$_{BL}$ lines. The programmed state of the selected cell can be read by sourcing a voltage to the read$_{WL}$ and sensing the current at the read$_{BL}$ while forcing it at ground, or vice versa.

\textbf{TD-VMM analog compute accuracy in terms of ENOB:}
In order to experimentally verify the VMM accuracy performance, we have measured the native $16\times16$ weights of our TD-VMM cores (weight extraction method detailed in Appendix - A), with a typical chip weight map reported in Fig.~\ref{fig2}c.

After the extraction, a set of $512\times16$ pulse-width input vectors (inset of Fig.~\ref{fig2}c) has been sent to the chip. The analog core, by processing the given inputs, has executed the 512 $16\times16$ TD-VMM operations, and the histogram of the resulting $\Delta V_{OUT}$ curves is compared in Fig.~\ref{fig2}d with the one constructed from theoretically expected results. The same experimental and theoretical data are used to obtain the scatter plot in Fig.~\ref{fig2}e, where each color refers to a different VMM BL column. The histogram of Fig.~\ref{fig2}f is built considering the errors related to all the 16 BL outputs, from which we were able to extract an overall RMS error (defined as $\Delta V_{\rm OUT,EXP}-\Delta V_{\rm OUT,REF}$) of 10.21~mV. Thus, by considering the RMS $\Delta V_{\rm{OUT,REF}}$ value of 648.2~mV, we extract an aggregate ENOB of the VMM of 5.7 bits [$\rm{ENOB} = (\rm{SNDR}_{dB}-1.76)/6.02$], where $\rm{SNDR}$ is here calculated as the ratio of the $\Delta V_{\rm OUT,REF}$ RMS value to the RMS error. The same approach has been used in Fig.~\ref{fig2}g to compute the ENOB for each independent BL column of the TD-VMM, from three separate dies, with all of them showing a typical value of the ENOB close to 5.7 bits.


\subsection{Temperature dependence of experimentally programmed weights}


In this section we analyze in detail the temperature impact on programmed weights. Due to the sub-threshold operation of 1T-FG cells, experimental weights strongly depend on the operating temperature. We try to explain this relation with three main effects, which are: (1) a weight magnification when varying the temperature; (2) a bending of the weight distribution curvature (which specifically depends on the target weight); (3) and distribution spread, which does not depend on the target weight.
Ideally, a pure weight (de)magnification applied to the whole network layer parameters would not have direct impact on the weight ENOB, and would not impact the classification accuracy of our classifier. However, in the hardware implementation, for a fixed set of input data, the weight magnification with increasing temperature moves the corresponding outputs in a voltage region where the precision of the VMM is degraded, for instance towards the charge amplifier saturation. This issue could be compensated by rescaling the input data, as long as the pulse-width rescaling does not push the TD-VMMs beyond their bandwidth limits. On the other hand, both bending and spread have a direct impact on the weight ENOB, as the actual temperature-related magnification of each weight depends on its corresponding nominal weight value  (resulting in distortion), and weights programmed to the same target value at a nominal temperature, spread out when $T$ is changed. 
Adapting a dynamic read voltage can compensate for both magnification and distortion, exploiting the fact that the cells are biased in sub-threshold region. Indeed, the read voltage operates on the weights in the opposite direction with respect to the temperature, i.e. if $V_{\rm READ}$ is reduced, we can observe both a weight compression and a curvature of the scatter plot, that can contrast the magnification and bending induced by the temperature raise. In Fig.~\ref{fig_ex_1} and Fig.~\ref{fig_ex_2} we report the 1T-FG experimentally programmed unsigned weights for the hidden- and output-layer, respectively. Scatter plots of experimental weights plotted against the respective theoretical target are reported for different $V_{\rm READ}$ and temperature conditions. From these figures, we are able to generate the plots for weight magnification and ENOB of the hidden+output layers unsigned weights of Fig.~\ref{fig3}j and k, respectively. The corresponding 2T-$W_{EXP}$ dependence on the temperature, when a constant $V_{\rm READ}$ is used, is shown in Fig.~\ref{fig3}i. After exploiting a $V_{\rm READ} = 1.15~\rm{V} - 3~\rm{mV/^{\circ}C} \cdot (T - 30^{\circ}\rm{C})$ correction rule, such dependence is strongly reduced, as reported in Fig.~\ref{fig3}l.

\begin{figure*}[t!]%
\centering
\includegraphics[width=0.82\textwidth]{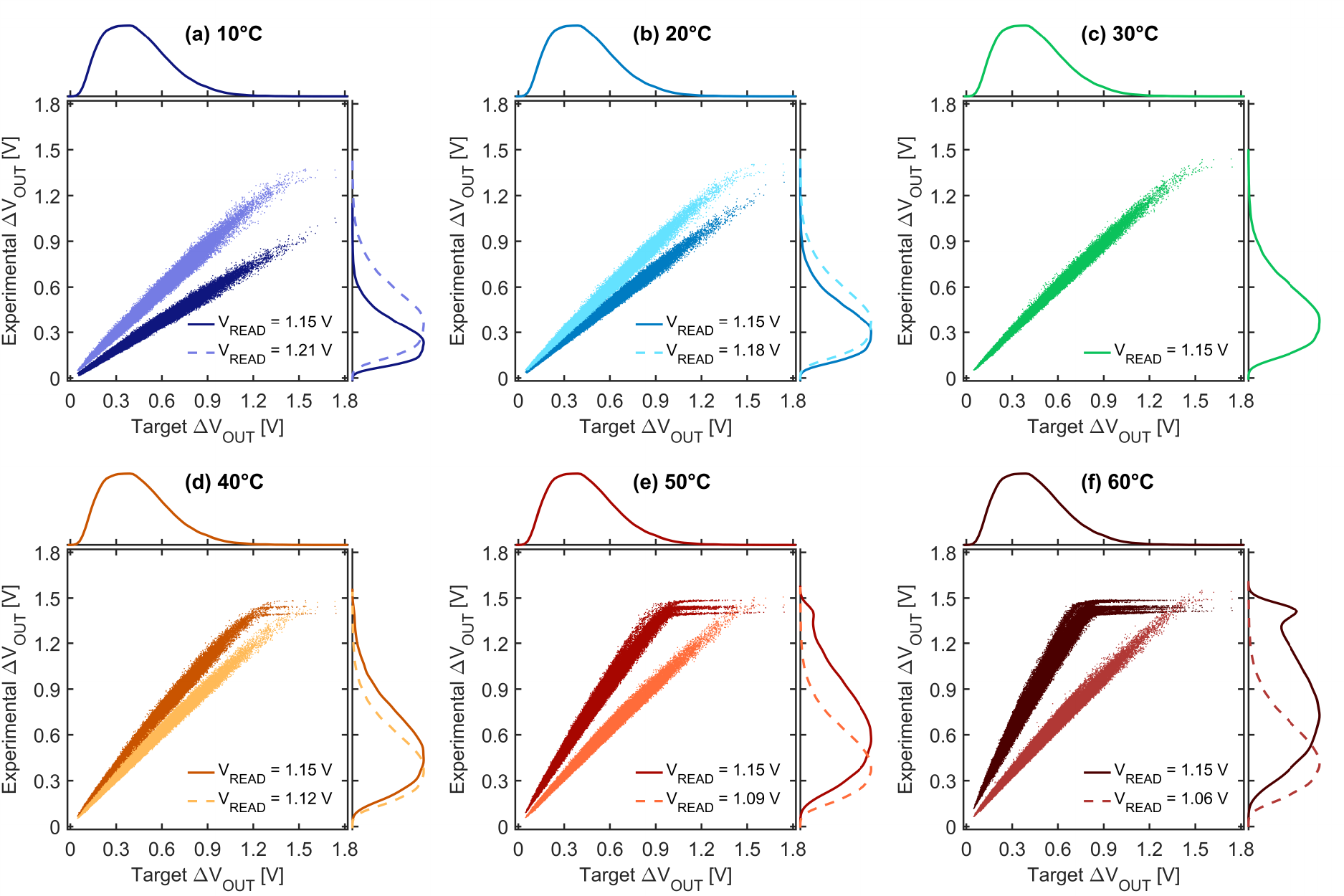}
\caption{\textbf{Experimental hidden-layer VMM output scatter plots and distributions.} 
Scatter plot of the experimental $\Delta V_{\rm OUT}$ plotted against the respective target (obtained through software simulation) for temperature from $10^{\circ}$C \textbf{(a)} to $60^{\circ}$C \textbf{(f)}. Two options are reported for each temperature: constant $V_{\rm{READ}} = 1.15~\rm{V}$ case, showing the drift of the distributions towards low levels of the output range (low temperature) or towards the output range saturation (high temperature); $V_{\rm READ}$ correction case ($\Delta V_{\rm READ} / \Delta T =  - 3~\rm{mV/^{\circ}C}$) showing the obtained resilience of the output distributions against the temperature variations. Reported  data points have been computed by the hardware VMM by processing 8000 low-resolution MNIST test-images.
}
\label{fig_ex_3}

\includegraphics[width=0.82\textwidth]{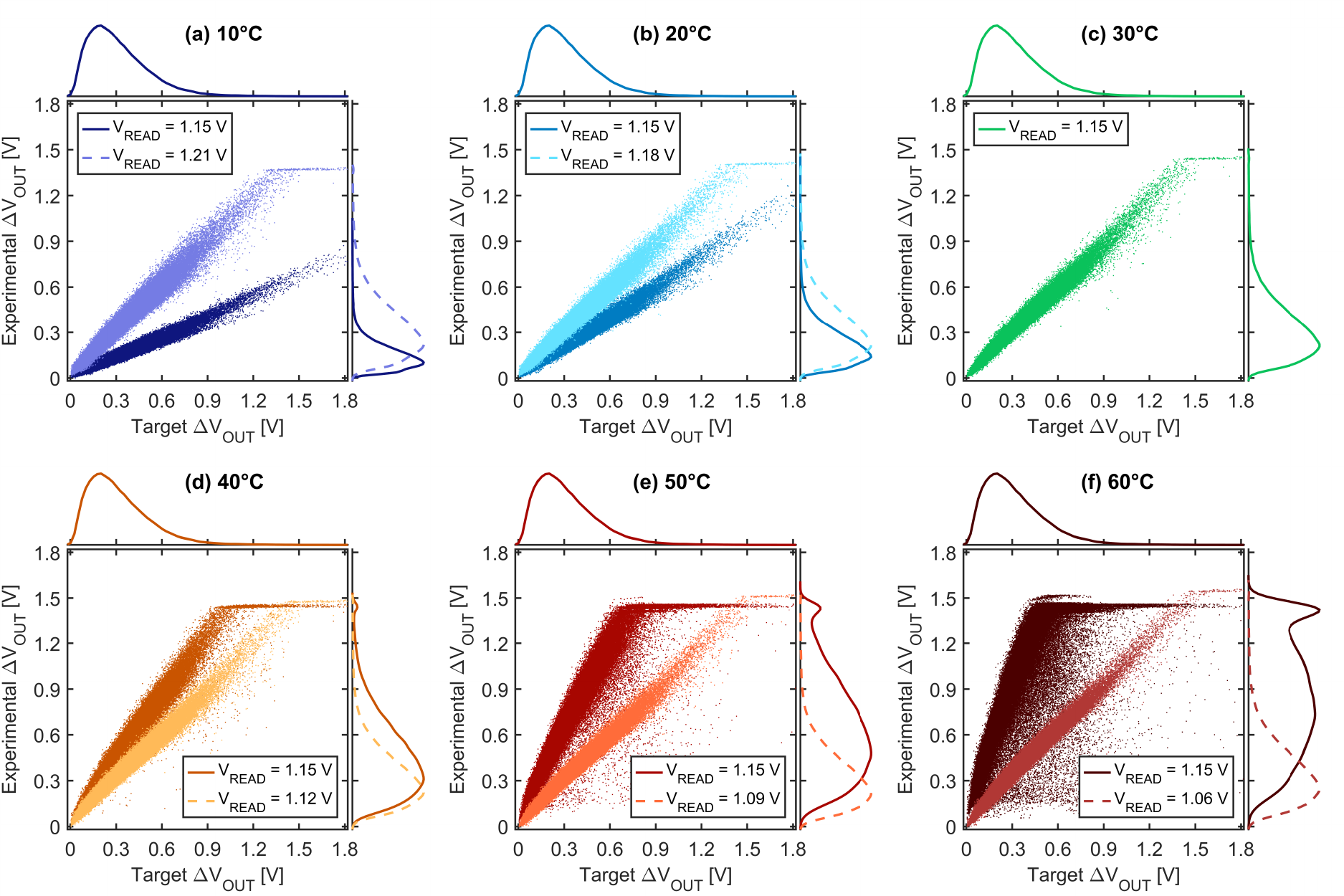}
\caption{\textbf{Experimental output-layer VMM output scatter plots and distributions.} 
Scatter plot of the experimental $\Delta V_{\rm OUT}$ plotted against the respective target (obtained through software simulation) for temperatures from $10^{\circ}$C \textbf{(a)} to $60^{\circ}$C \textbf{(f)}. Two options are reported for each temperature: constant $V_{\rm{READ}} = 1.15~\rm{V}$ case, showing the drift of the distributions towards low levels of the output range (low temperature) or towards the output range saturation (high temperature); $V_{\rm{READ}}$ correction case ($\Delta V_{\rm{READ}} / \Delta T =  -3~\rm{mV/^{\circ}C}$) showing the obtained resilience of the output distributions against the temperature variations. Reported  data points have been computed by the hardware VMM by processing 8000 low-resolution MNIST test-images.
}
\label{fig_ex_4}
\end{figure*}

\subsection{Low resolution MNIST test: TD-VMM outputs}

The low-resolution (4$\times$4) MNIST test set presented in Fig.~\ref{fig1}, with digits from `1' to `8', has been executed to characterize the full neural network chip, as summarized in Fig.~\ref{fig4}. To this purpose, data reported in Fig.~\ref{fig4}e-f (normalized histograms of the output-layer TD-VMM outputs as obtained for varying temperature and at constant or varying $V_{\rm READ}$), have been extrapolated from Fig.~\ref{fig_ex_3} and Fig.~\ref{fig_ex_4}. In these figures, we report the outputs of the TD-VMM cores when processing the low resolution MNIST test at the hidden- and output-layer, respectively. The scatter plots have been constructed by plotting the VMM outputs obtained for all 8000 test images. The test on the full dataset has been repeated for 6 temperatures, from 10$^{\circ}$C up to 60$^{\circ}$C (10$^{\circ}$C step), by considering a constant $V_{\rm READ}$ read mode (dark data points) and a $V_{\rm READ}$ linear temperature correction method (light data point).

\begin{IEEEbiography}[{\includegraphics[width=1in,height=1.25in,clip,keepaspectratio]{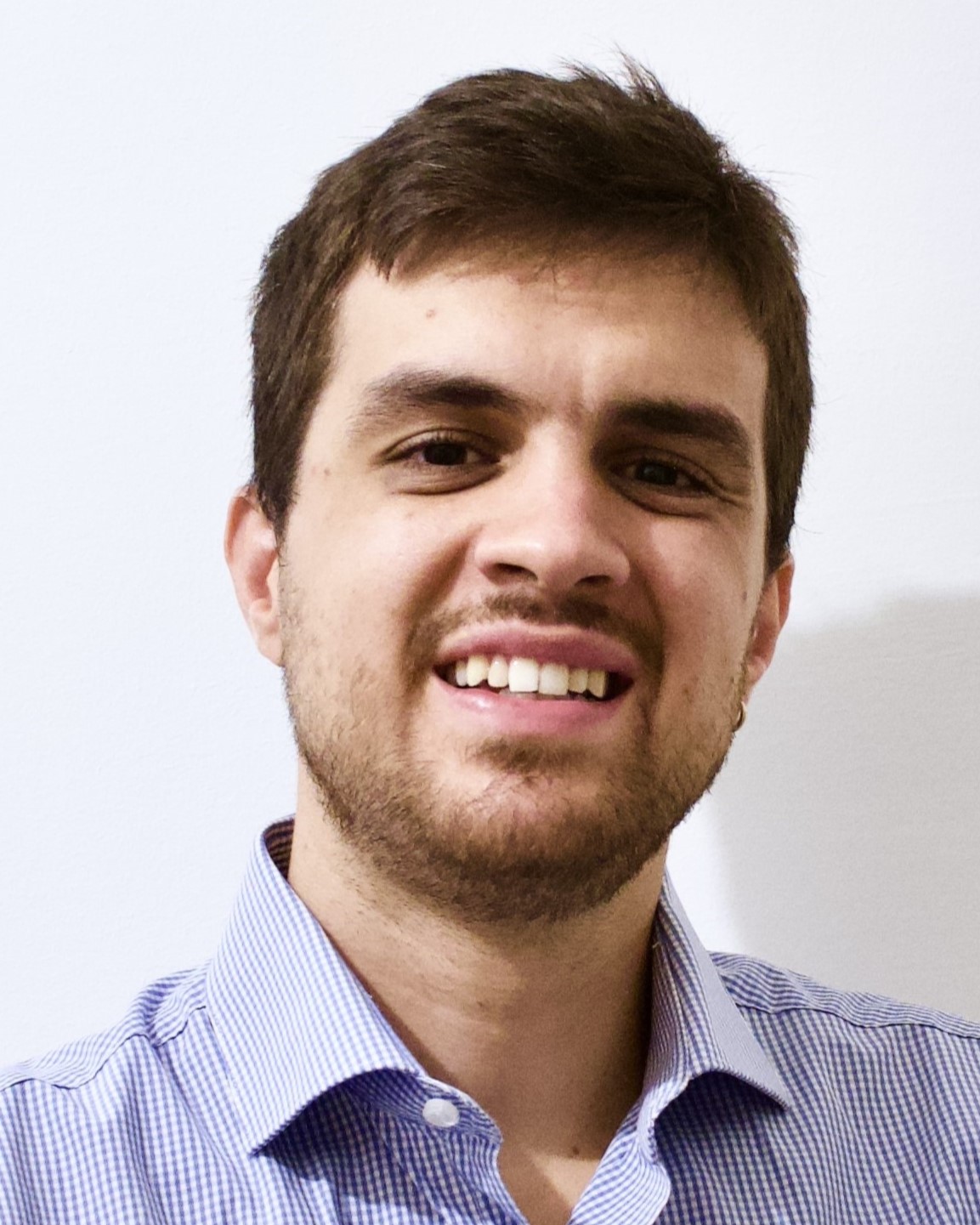}}]{Tommaso Rizzo}
received the M.S. degree (2019), and the Ph.D. degree (2023) in EE from the University of Pisa, both cum laude. He obtained his Ph.D. in a joint program with Quantavis s.r.l., Pisa, working as a research engineer, with a thesis on analog IC design for implantable neuromorphic devices. From 2014 to 2019, he was “Allievo Ordinario” at Sant’Anna School of Advanced Studies, Pisa. In 2017, he was with Fermilab, Batavia, IL, USA, and in 2019, he joined imec, Eindhoven, NL, both as a visiting student. He is currently with STMicroelectronics, Pisa, working in the advanced analog R\&D product division. His research focuses on advanced analog IC design using standard and non-standard CMOS technologies. 
\end{IEEEbiography}

\begin{IEEEbiography}[{\includegraphics[width=1in,height=1.25in,clip,keepaspectratio]{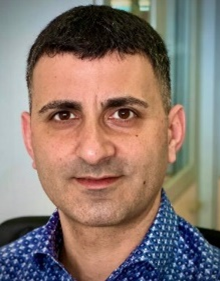}}]{Sebastiano Strangio}
received the B.S. and M.S. degrees (cum laude) in EE, and the Ph.D. degree from the University of Calabria, Cosenza, Italy, in 2010, 2012, and 2016, respectively. 
He was with IMEC, Leuven, Belgium, as a Visiting Student, in 2012, working on the electrical characterization of resistive-RAM memory cells. 
From 2013 to 2016, he was a Temporary Research Associate with the University of Udine, and with the Forschungszentrum Jülich, Germany, as a Visiting Researcher, in 2015, researching on TCAD simulations, design, and characterization of TFET-based circuits. 
From 2016 to 2019, he was with LFoundry, Avezzano, Italy, where he worked as a Research and Development Process Integration and Device/TCAD Engineer, with main focus on the development of a CMOS Image Sensor Technology Platform. 
He is currently a Researcher in electronics with the University of Pisa. 
He has authored or coauthored over 40 papers, most of them published in IEEE journals and conference proceedings. 
His research interests include technologies for innovative devices (e.g. TFETs) and circuits for innovative applications (CMOS analog building blocks for DNNs), as well as CMOS image sensors, power devices and circuits based on wide-bandgap materials. 
\end{IEEEbiography}

\begin{IEEEbiography}[{\includegraphics[width=1in,height=1.25in,clip,keepaspectratio]{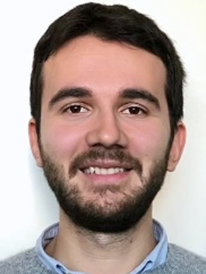}}]{Alessandro Catania}
received the B.S., M.S., and Ph.D. degrees in electronic engineering from the University of Pisa, Italy, in 2014, 2016 and 2020, respectively. 
He is currently working as an Assistant Professor with the University of Pisa. 
His current research interests include mixed-signal microelectronic design for harsh environments and wireless power transfer systems for implantable systems. 
\end{IEEEbiography}

\begin{IEEEbiography}[{\includegraphics[width=1in,height=1.25in,clip,keepaspectratio]{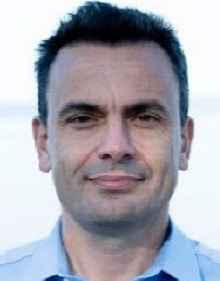}}]{Giuseppe Iannaccone}
received the M.S. and Ph.D. degrees in EE from the University of Pisa, in 1992 and 1996, respectively. 
He is currently Deputy President and Professor of electronics with the University of Pisa. 
He has coordinated several European and national projects involving multiple partners and has acted as principal investigator in several research projects funded by public agencies at the European and national level, and by private organizations. 
He co-founded the academic spinoff Quantavis s.r.l. and is involved in other technology transfer initiatives. 
He has authored or coauthored more than 250 articles published in peer-reviewed journals and more than 160 papers in proceedings of international conferences, gathering more than 12000 citations on the Scopus database. 
His research interests include quantum transport and noise in nanoelectronic and mesoscopic devices, development of device modeling tools, new device concepts and circuits beyond CMOS technology for artificial intelligence, cybersecurity, implantable biomedical sensors, and the Internet of Things. 
He is a fellow of the American Physical Society. 
\end{IEEEbiography}

\end{document}